\documentclass[epj]{svjour} 
\usepackage{graphicx,amssymb}
\newcommand{\dgs}{$^\circ$}
\def\nuc#1#2{\relax\ifmmode{}^{#1}{\protect\text{#2}}\else${}^{#1}$#2\fi}
\newcommand{\infnpd}{1}
\newcommand{\infnge}{2}
\newcommand{\infnmi}{3}
\newcommand{\lisboa}{4}
\newcommand{\lngs}{5}
\newcommand{\atomki}{6}
\newcommand{\infnto}{7}
\newcommand{\unina}{8}
\newcommand{\lnl}{9}
\newcommand{\caserta}{10}
\newcommand{\bochum}{11}
\newcommand{\teramo}{12}
\newcommand{\berlin}{13}

\begin{document}

% =========================================================
\title{Feasibility of low energy radiative capture experiments at the LUNA underground accelerator facility}
\author{
	D.\,Bemmerer \inst{\infnpd,\berlin} 
		\thanks{\emph{E-mail address:} {\tt bemmerer@pd.infn.it}} \and
	F.\,Confortola \inst{\infnge}  \and
	A.\,Lemut \inst{\infnge} \and
	R.\,Bonetti \inst{\infnmi} \and
	C.\,Broggini \inst{\infnpd} \and
	P.\,Corvisiero \inst{\infnge} \and
	H.\,Costantini \inst{\infnge} \and
	J.\,Cruz \inst{\lisboa} \and
	A.\,Formicola \inst{\lngs} \and
	Zs.\,F\"ul\"op \inst{\atomki} \and
	G.\,Gervino \inst{\infnto} \and
	A.\,Guglielmetti \inst{\infnmi} \and
	C.\,Gustavino \inst{\lngs} \and
	Gy.\,Gy\"urky \inst{\atomki} \and
	G.\,Imbriani \inst{\unina} \and
	A.P.\,Jesus \inst{\lisboa} \and
	M.\,Junker \inst{\lngs} \and
	B.\,Limata \inst{\unina} \and
	R.\,Menegazzo \inst{\infnpd,\lnl} \and
	P.\,Prati \inst{\infnge} \and
	V.\,Roca \inst{\unina} \and
	D.\,Rogalla \inst{\caserta} \and
	C.\,Rolfs \inst{\bochum} \and
	M.\,Romano \inst{\unina} \and
	C.\,Rossi Alvarez \inst{\infnpd} \and
	F.\,Sch\"umann \inst{\bochum} \and
	E.\,Somorjai \inst{\atomki} \and
	O.\,Straniero \inst{\teramo} \and
	F.\,Strieder \inst{\bochum} \and
	F.\,Terrasi \inst{\caserta} \and
	H.P.\,Trautvetter \inst{\bochum} \and
	A.\,Vomiero \inst{\lnl}
	}% author
%
%\offprints{D. Bemmerer}          % Insert a name or remove this line
%
\institute{
	INFN, Sezione di Padova, via Marzolo 8, 35131 Padova, Italy % 1
	\and 
	Dipartimento di Fisica, Universit\`a di Genova, and INFN, Genova, Italy % 2
	\and 
	Istituto di Fisica, Universit\`a di Milano, and INFN, Milano, Italy % 3
	\and 
	Centro de Fisica Nuclear da Universidade de Lisboa, Lisboa, Portugal  % 4
	\and 
	INFN, Laboratori Nazionali del Gran Sasso, Assergi, Italy % 5
	\and 
	ATOMKI, Debrecen, Hungary % 6
	\and
	Dipartimento di Fisica Sperimentale, Universit\`a di Torino, and INFN, Torino, Italy % 7
	\and
	Dipartimento di Scienze Fisiche, Universit\`a di Napoli "Federico II", and INFN, Sezione di Napoli, Napoli, Italy % 8
	\and
	INFN, Laboratori Nazionali di Legnaro, Legnaro, Italy % 9
	\and
	Seconda Universit\`a di Napoli, Caserta, and INFN, Sezione di Napoli, Napoli, Italy % 10
	\and
	Institut f\"ur Experimentalphysik III, Ruhr-Universit\"at Bochum, Bochum, Germany % 11
	\and
	Osservatorio Astronomico di Collurania, Teramo, and INFN, Sezione di Napoli, Napoli, Italy % 12
	\and
	Institut f\"ur Atomare Physik und Fachdidaktik, Technische Universit\"at Berlin, Berlin, Germany % 13
	}% institute
\date{Received: date / Revised version: date}

% =========================================================
\abstract{
The LUNA (Laboratory Underground for Nuclear Astrophysics) facility has been designed to study nuclear reactions of astrophysical interest.
It is located deep underground in the Gran Sasso National Laboratory, Italy. 
Two electrostatic accelerators, with 50 and 400\,kV maximum voltage, in combination with solid and gas target setups allowed to measure the total cross sections of the radiative capture reactions \nuc{2}{H}(p,$\gamma$)\nuc{3}{He} and \nuc{14}{N}(p,$\gamma$)\nuc{15}{O} within their relevant Gamow peaks. 
We report on the gamma background in the Gran Sasso laboratory measured by germanium and bismuth germanate detectors, with and without an incident proton beam.
A method to localize the sources of beam induced background using the Doppler shift of emitted gamma rays is presented.
The feasibility of radiative capture studies at energies of astrophysical interest is discussed for several experimental scenarios.
\PACS{
	{25.40.Lw}{Radiative capture} \and
	{26.20.+f}{Hydrostatic stellar nucleosynthesis} \and
	{29.17.+w}{Electrostatic, collective, and linear accelerators} \and
	{29.30.Kv}{X- and gamma-ray spectroscopy}
}% PACS
}% abstract
% =========================================================
\authorrunning{D.\,Bemmerer, F.\,Confortola, A.\,Lemut {\it et al.}}
\titlerunning{Feasibility of low energy radiative capture experiments at the LUNA...}

% =========================================================
\maketitle

% =========================================================
\section{Introduction}
\label{Introduction}
% =========================================================

Stars generate energy and synthesize chemical elements in thermonuclear reactions \cite{Eddington20,Bethe39,Rolfs88}. All reactions induced by charged particles in a star take place in an energy window called the Gamow peak. 
For the \nuc{14}{N}(p,$\gamma$)\nuc{15}{O} reaction, to give an example, the Gamow peak lies between 20 and 80\,keV for a star in a globular cluster which is at the evolution stage used for the cluster age determination.
\par 

The cross section $\sigma(E)$ of a charged particle induced reaction drops steeply with decreasing energy due to the Coulomb barrier in the entrance channel:
\begin{equation} \label{S-factor}
\sigma(E) = \frac{S(E)}{E} e^{-2 \pi \eta} 
\end{equation}
where $S(E)$ is the astrophysical S factor \cite{Rolfs88}, and $\eta$ is 
the Sommerfeld parameter with $2 \pi \eta$\,=\,$31.29\ Z_1 Z_2 \sqrt{\frac{\mu}{E}}$. Here $Z_1$ and $Z_2$ are the charge numbers of projectile and target nucleus, respectively, $\mu$ is the reduced mass (in amu units), and $E$ is the center of mass energy (in keV units).\par

In the static burning of stars, $\sigma(E)$ has a very low value at the Gamow peak. This prevents a direct measurement in a laboratory at the earth's surface, where the signal to background ratio is too small because of cosmic ray interactions. Hence, cross sections are measured at high energies and expressed as the astrophysical S factor from eq. (\ref{S-factor}). The S factor is then used to extrapolate the data to the relevant Gamow peak. Although $S(E)$ varies only slowly with energy for the direct process, resonances and resonance tails may hinder an extrapolation, resulting in large uncertainties \cite{Rolfs88}. Therefore, the primary goal of experimental nuclear astrophysics remains to measure the cross section at energies inside the Gamow peak, or at least to approach it as closely as possible.\par

The Laboratory Underground for Nuclear Astrophysics (LUNA) has been designed for this purpose and is located in the Laboratori Nazionali del Gran Sasso (LNGS) in Italy. LUNA uses high current accelerators with small energy spread in combination with high efficiency detection systems, which are described below.\par

At the 50\,kV LUNA1 accelerator \cite{Greife94}, the \linebreak 
\nuc{3}{He}(\nuc{3}{He},2p)\nuc{4}{He} cross section was measured for the first time  within its solar Gamow peak \cite{Junker98,Bonetti99}. Subsequently, a windowless gas target setup and a $4\pi$ bismuth germanate (BGO) summing detector \cite{Casella02-NIMA} have been used to study the radiative capture reaction \nuc{2}{H}(p,$\gamma$)\nuc{3}{He}, also within its solar Gamow peak \cite{Casella02-NuclPhysA}.\par

The 400\,kV LUNA2 accelerator \cite{Formicola03-NIMA} has been used to study the radiative capture reaction \nuc{14}{N}(p,$\gamma$)\nuc{15}{O}, which is the bottleneck of the hydrogen burning CNO cycle \cite{Bethe39}. Most previous experiments on this reaction \cite[and references therein]{Schroeder87} had the lowest yield point at $E$\,=\,240\,keV, much higher than the Gamow peak. For the LUNA\linebreak
\nuc{14}{N}(p,$\gamma$)\nuc{15}{O} study, titanium nitride (TiN) solid 
targets and a high purity germanium detector were used to measure the cross sections for the transitions to several states in \nuc{15}{O}, including the ground state, down to $E$\,=\,130\,keV \cite{Formicola03-Debrecen,Formicola04-PhysLettB,Formicola04-Diss}. The LUNA data resulted in a total extrapolated S factor that was a factor 2 smaller than the values adopted by recent compilations \cite{Adelberger98,NACRE99}, leading to considerable astrophysical consequences \cite{Bahcall04,Imbriani04,Innocenti04}. In order to extend the \nuc{14}{N}(p,$\gamma$)\nuc{15}{O} cross section data to even lower energies, a gas target setup similar to the one used for the \nuc{2}{H}(p,$\gamma$)\nuc{3}{He} study and a BGO detector have been installed at the LUNA2 400\,kV accelerator \cite{LUNA-AR2003}. \par

In the present work, the features of the LUNA facility are reviewed. A solid target setup and a gas target setup, both for the study of radiative capture reactions, and a setup specifically designed for background studies are described. The $\gamma$ background relevant to radiative capture experiments
is discussed for each setup, with and without a proton beam incident on the target. A procedure to localize the source of ion beam induced background using the Doppler shift of emitted $\gamma$ rays is employed.
The feasibility of low energy radiative capture experiments at the LUNA facility is evaluated.
\par

%%%%%%%%%%%%%%
\section{The Gran Sasso underground laboratory} 
%%%%%%%%%%%%%%

The Gran Sasso underground laboratory\footnote{Web page: \tt http://www.lngs.infn.it} consists of three experimental halls and several connecting tunnels. Its site is protected from cosmic rays by a rock cover equivalent to 3800\,m water (3800\,m\,w.e.). The LUNA facility is situated in a bypass tunnel, about 30\,m to the west of the entrance of experimental hall A. \par

The overlying rock suppresses the flux of cosmic ray induced muons by six orders of magnitude \cite{Ahlen90}, resulting in a flux of muon induced neutrons of the order of $\Phi_{\rm n_\mu}$\,$\approx$\,$10^{-8} \rm\frac{n}{cm^2 \cdot s}$, according to a recent simulation \cite{Wul04}. 
The measured total neutron flux in hall A is somewhat higher, $\Phi_n$\,$\approx$\,$4 \cdot 10^{-6} \rm\frac{n}{cm^2\ s}$ \cite{Belli89}. 
This excess can be explained with neutrons from ($\alpha$,n) reactions and spontaneous fission of \nuc{238}{U}, both of which take place in the surrounding rock and the concrete walls of the experimental areas \cite{Wul03}. The neutron flux data for hall A offer an approximate picture for the situation at the LUNA site, since the rock and concrete surroundings are comparable.\par

In a previous experiment using germanium detectors, the laboratory background was compared between a facility at sea level and an underground site shielded by 500\,m\,w.e. \cite{Wordel96}. It was shown that for $E_\gamma$\,$>2$\,MeV, the counting rate at sea level was dominated by cosmic rays, especially muons, traversing the detector. For $E_\gamma$\,$\leq 2$\,MeV, a 15\,m\,w.e. cosmic ray shield achieved a sizable reduction in both line and continuum background, mainly by reducing the flux of cosmic ray induced neutrons \cite{Heusser93}. At LNGS, a reduction in the $\gamma$ continuum of about a factor 100 was observed \cite{Arpesella96} for the same energy region when compared to a low level counting facility at the earth's surface.\par

These previous studies focused on the energy region $E_\gamma$\,$<$\,3\,MeV relevant to activity measurements. Radiative capture reactions \cite{Fiorentini95} often lead to the emission of $\gamma$ rays of higher energy. Since direct and indirect effects of cosmic rays dominate the counting rate at high $\gamma$ energies, active shielding with a muon detector is generally used to suppress this background in laboratories at the earth's surface. 
An active muon shield can reduce the background counting rate by about a factor 10\,-\,50 for $E_\gamma$\,=\,7\,-\,11\,MeV \cite{Mueller90}. 
The $10^{-6}$ reduction in cosmic ray induced muons provided by the Gran Sasso rock cover therefore offers a clear advantage, especially at high $\gamma$ energies.\par

%%%%%%%%%%%%%%
\section{LUNA setups to study radiative capture reactions}
%%%%%%%%%%%%%%

Radiative proton capture experiments were carried out at the LUNA2 400\,kV accelerator \cite{Formicola03-NIMA}. Three different target systems called setup A, B, and C were used; they are described below.

%%%%%%%%%%%%%%
\subsection*{Setup A: Solid target and germanium detector}
%%%%%%%%%%%%%%

Setup A consists of a TiN solid target and a high purity germanium detector in close geometry at 55\dgs\ angle to the beam direction, with the detector endcap at 1.5\,cm distance from the target. It is similar to a setup described elsewhere \cite{Formicola03-NIMA}. For the purpose of the present work, only spectra without beam, taken with a p type germanium detector of 108\,\% relative efficiency, are used. \par

%%%%%%%%%%%%%%
\subsection*{Setup B: Windowless gas target and $4\pi$ BGO detector}
%%%%%%%%%%%%%%

\begin{figure}[bt]
 \centering 
 \includegraphics[width=85mm]{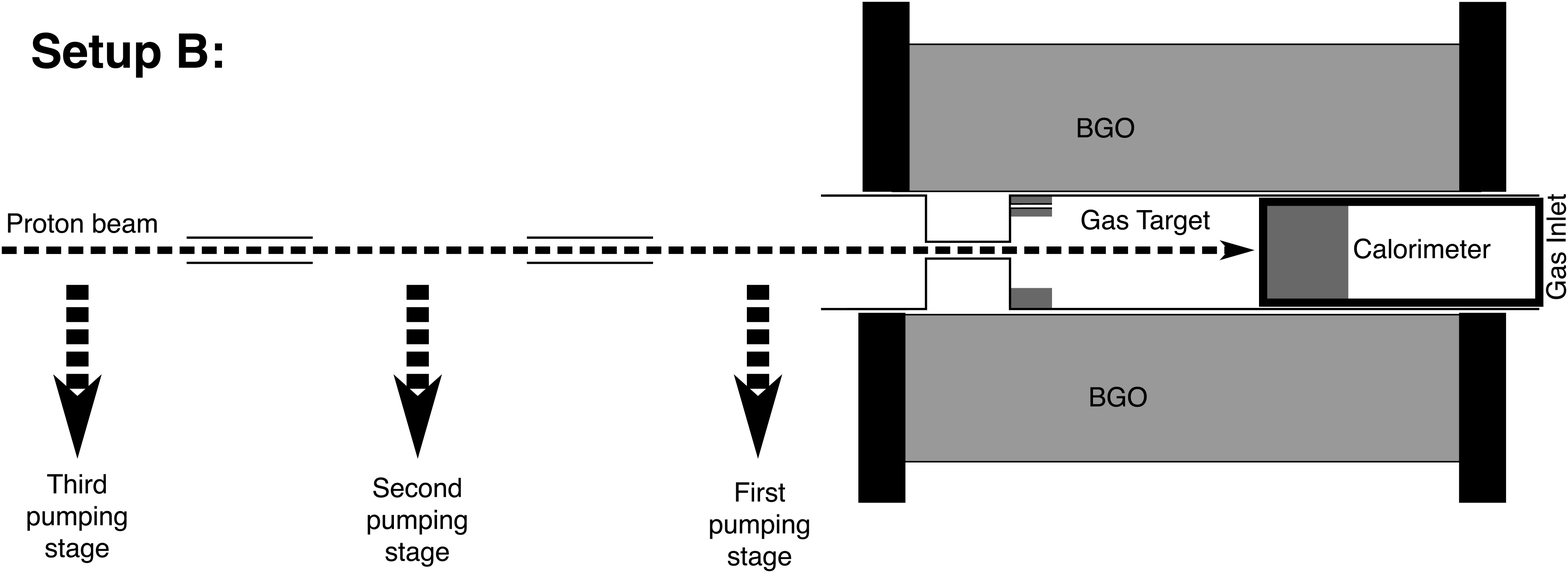}
 \includegraphics[width=85mm]{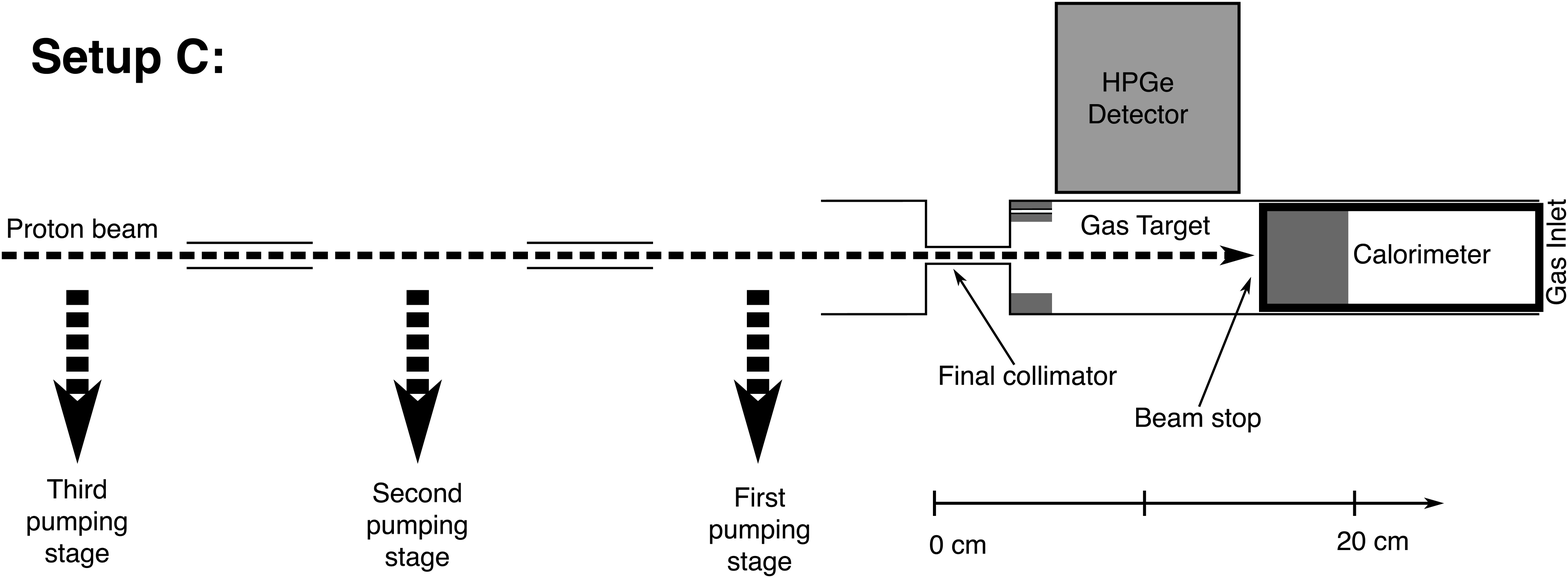}
 \caption{Sketch of setups B (top) and C (bottom). The sizes of the target and detectors are to scale, while the rest of the setup is shown schematically.}
 \label{Setup-B.pdf}
\end{figure}

A sketch of setup B is shown at the top of fig. \ref{Setup-B.pdf}. It is a modified version of the LUNA \nuc{2}{H}(p,$\gamma$)\nuc{3}{He} setup
\cite{Casella02-NIMA}, with a 12\,cm long target cell. In the experiment, a proton beam of energy $E_{\rm p}$\,=100\,-\,400\,keV and current up to 500\,$\mu$A is supplied by the LUNA2 400\,kV accelerator and enters the three stage, differentially pumped windowless gas target system through a succession of water cooled apertures; the final aperture has a diameter of 7\,mm and a length of 40\,mm
%Daniel 15 Jan 05
and is made from brass.
%Daniel 15 Jan 05
The target cell is fitted into the 6\,cm wide bore hole at the center of an annular BGO detector having 7\,cm radial thickness and 28\,cm length. Also inside the BGO bore hole is a calorimeter for the measurement of the beam intensity, with a 41\,mm thick block of oxygen free copper serving as the beam stop.\par

% ============ Setup-B.pdf and Setup-C.pdf

The target gases were 0.5\,-\,2.0\,mbar nitrogen 
of chemical purity 99.9995\,\%, 
0.5\,-\,2.0\,mbar helium 
of chemical purity 99.9999\,\%, 
and an empty target cell ($< 10^{-3}$\,mbar).
The first pumping stage is evacuated by a WS\,2000 roots blower, leading to a pressure ratio between target and first pumping stage that is better than a factor 100. The second and third pumping stages are at $10^{-5}$ and $10^{-6}$\,mbar pressure, respectively. The BGO detector has an absolute peak detection efficiency of 65\,-\,70\,\% for 3\,-\,10\,MeV $\gamma$ rays emitted from a point-like source at the center of the detector bore hole \cite{Casella02-NIMA}. \par 

%%%%%%%%%%%%%%
\subsection*{Setup C: Windowless gas target and germanium detector}
%%%%%%%%%%%%%%

Because of the poor energy resolution of BGO detectors, the summing crystal of setup B gives only limited spectroscopic information. In order to reliably identify and localize the sources of beam induced background, the modified setup C is used (bottom of fig. \ref{Setup-B.pdf}).
A high purity germanium detector with 120\,\% relative efficiency is placed close to the target chamber, at an angle of 90\dgs\ to the beam direction. \par

%%%%%%%%%%%%%%
\section{Background in the LUNA setups}
%%%%%%%%%%%%%%

In order to evaluate the feasibility of radiative capture experiments at the three different setups, various background conditions have to be investigated.
In the present section, the laboratory background and the background induced by a proton beam incident on the target system are discussed, and the Doppler shift of $\gamma$ rays from ion beam induced background is used to localize their source.\par

%%%%%%%%%%%%%%
\subsection{Laboratory background}
%%%%%%%%%%%%%%

\begin{figure}[tb]
 \centering 
 \includegraphics[width=85mm]{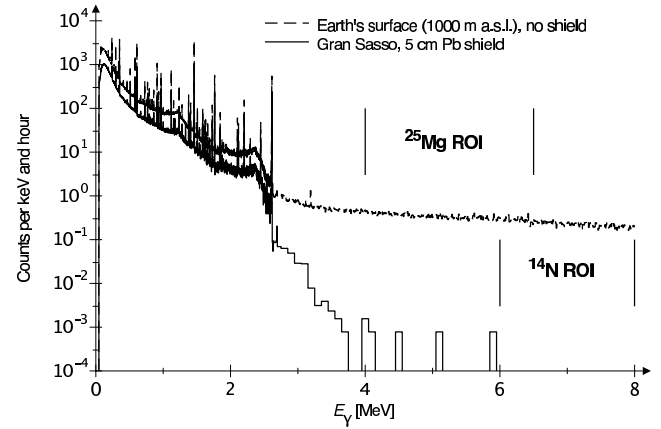}
 \caption{Laboratory $\gamma$ background as seen with the germanium detector of setup A at the earth's surface (1000 m above sea level) and inside the Gran Sasso underground facility. }
 \label{GeBochum-inout.pdf}
\end{figure}

The laboratory background taken with the germanium detector of setup A is shown in fig. \ref{GeBochum-inout.pdf}, where spectra recorded at the earth's surface and inside the Gran Sasso laboratory are compared. Above the 2.61\,MeV line from \nuc{208}{Tl}, the surface spectrum has been rebinned in 10\,keV bins, and the underground spectrum has been rebinned in \linebreak 100\,keV bins. In the plot, possible regions of interest (ROI) for the LUNA \nuc{14}{N}(p,$\gamma$)\nuc{15}{O} ($Q$\,=\,7.30\,MeV) experiment and a possible \nuc{25}{Mg}(p,$\gamma$)\nuc{26}{Al} ($Q$\,=\,6.31\,MeV) study are marked. One can see that the Gran Sasso mountain effectively eliminates the muon induced continuum which dominates the surface spectrum above 2.6\,MeV. In the underground spectrum, the tail from 2.6 to 3.7\,MeV is due to coincidence summing events from \nuc{208}{Tl}; the remaining counts above 3.7\,MeV are caused by neutron capture, mainly in the germanium detector material.
\par

% ============ GeBochum-inout.pdf

\begin{figure}[tb]
 \centering 
 \includegraphics[width=85mm]{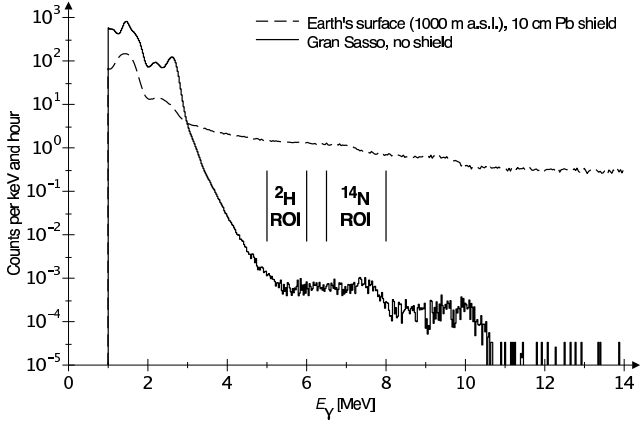}
 \caption{Same as fig. \ref{GeBochum-inout.pdf}, but for the BGO detector of setup B.}
 \label{NatBG-BGO-inout.pdf}
\end{figure}

Analogous spectra for the BGO detector of setup B are plotted in fig. \ref{NatBG-BGO-inout.pdf}, with the underground spectrum rebinned in 25\,keV bins.  Possible regions of interest for the LUNA \nuc{2}{H}(p,$\gamma$)\nuc{3}{He} ($Q$\,=\,5.49\,MeV) and \nuc{14}{N}(p,$\gamma$)\nuc{15}{O} experiments are marked in the figure. Up to $E_\gamma$\,=\,3.7\,MeV, both BGO spectra are dominated by natural radioisotopes, with the most prominent lines being the 1.46\,MeV \nuc{40}{K} line, a 2.20\,MeV \nuc{214}{Bi} line superimposed with a \linebreak 2.34\,MeV sum peak from the detector intrinsic contaminant \nuc{207}{Bi}, 
and the 2.61\,MeV \nuc{208}{Tl} line. The \nuc{40}{K}, \nuc{214}{Bi}, and \nuc{208}{Tl} lines from the room background can be reduced significantly using lead shielding, as can be seen by comparing the low energy parts of the spectra. %Daniel 15 Jan 05
This comparison also shows that possible \nuc{208}{Tl} impurities situated inside the BGO detector or at its surface contribute only weakly to the 2.61\,MeV counting rate. Therefore, no significant number of counts from the simultaneous detection of $\beta^-$ and $\gamma$ rays from \nuc{208}{Tl} $\beta^-$ decay ($Q_\beta$\,=\,5.00\,MeV) is expected.
%Daniel 15 Jan 05
\par

% ============ NatBG-BGO-inout.pdf

In the BGO spectrum taken at the earth's surface, the long plateau extending from 3.7\,MeV on upwards is caused by cosmic rays and, to a lesser degree, also muon induced neutrons. The barely recognizable shoulders at 7.5 and 11\,MeV correspond to the more visible shoulders in the spectrum taken underground, which are explained below.\par

% ================== 
\begin{table}[tb]
\centering
\caption{Counting rate without beam in counts per hour and keV for the standard 108\,\% germanium detector of setup A. Only statistical uncertainties are quoted.}
\label{NaturalBG-Countingrate-Ge-table}
\begin{tabular}{|l|c|c|}
\hline
Reaction \raisebox{3mm}{} & \nuc{25}{Mg}(p,$\gamma$)\nuc{26}{Al} & \nuc{14}{N}(p,$\gamma$)\nuc{15}{O} \\ 
Relevant $\gamma$ energy & 6.2\,MeV & 6.8\,MeV \\ \hline 
Earth's surface, & & \\
no shield, no beam & $0.262 \pm 0.002$ &  $0.221 \pm 0.002$ \\ 
Gran Sasso, & & \\
5\,cm Pb, no beam & $< 1.6 \cdot 10^{-4}\ (1\sigma)$ & $< 1.0 \cdot 10^{-4}\ (1\sigma)$ \\ 
\hline
\end{tabular}
\end{table}
% ================== 
\begin{table}[tb]
\centering
\caption{Same as table \ref{NaturalBG-Countingrate-Ge-table}, but for the BGO detector of setup B.}
\label{NaturalBG-Countingrate-BGO-table}
\begin{tabular}{|l|c|c|}
\hline
Reaction \raisebox{3mm}{} & \nuc{2}{H}(p,$\gamma$)\nuc{3}{He} & \nuc{14}{N}(p,$\gamma$)\nuc{15}{O} \\ 
$\gamma$ energy region & 5.0\,-\,6.0\,MeV & 6.5\,-\,8.0\,MeV \\ \hline
Earth's surface, & & \\
10\,cm Pb, no beam & 1.373 $\pm$ 0.007 & 0.959 $\pm$ 0.005  \\ 
Gran Sasso,  & & \\
no shield, no beam & $(7.1 \pm 0.2) \cdot 10^{-4}$ & $(6.1 \pm 0.2) \cdot 10^{-4}$  \\  
\hline
\end{tabular}
\end{table}
% ================== 

For the BGO spectrum taken underground, the long tail starting at 2.6\,MeV is caused by 
%Daniel 15 Jan 05 begin 
several 
%Daniel 15 Jan 05 end 
different phenomena. 
%Daniel 15 Jan 05 begin 
Up to 3.7\,MeV, there are unresolved natural radioisotope lines, mainly from the summing of two $\gamma$ rays from \nuc{208}{Tl} decay. 
Accidental coincidence between two \linebreak 2.61\,MeV \nuc{208}{Tl} $\gamma$ rays contributes counts at 5.2\,MeV; a rate of $10^{-3}$\,$\rm \frac{counts}{keV \cdot h}$ for this effect can be estimated from the 2.61\,MeV single counting rate. 
%Daniel 15 Jan 05 end 
The tail from 3.7 to 5.5\,MeV and the plateau from 5.5 to 8.0\,MeV are due to neutron capture in the BGO crystal, whose germanium content has several open channels for (n,$\gamma$) capture in that energy region, and to neutron capture in the copper calorimeter and the aluminium vacuum vessel. 
The plateau from 8.0 to 10.5\,MeV can be attributed to (n,$\gamma$) reactions on \nuc{54,57}{Fe} in the walls of the detector. For all these capture processes, both thermal and high energy neutrons contribute. 
Neutron capture on nuclides like \nuc{209}{Bi} often leads to a decay chain including $\alpha$ decays. However, $\alpha$ particles cannot contribute to the counting rate at high light outputs (corresponding to high $\gamma$ energies) because their light yield in a BGO scintillator is a factor 3\,-\,5 lower relative to $\gamma$ rays of the same energy \cite{Becchetti84,Dlouhy92}. The BGO spectrum shows few counts above 10.5\,MeV; they can be attributed to muons passing through parts of the detector. \par

\begin{figure}[bt]
 \centering 
 \includegraphics[width=85mm]{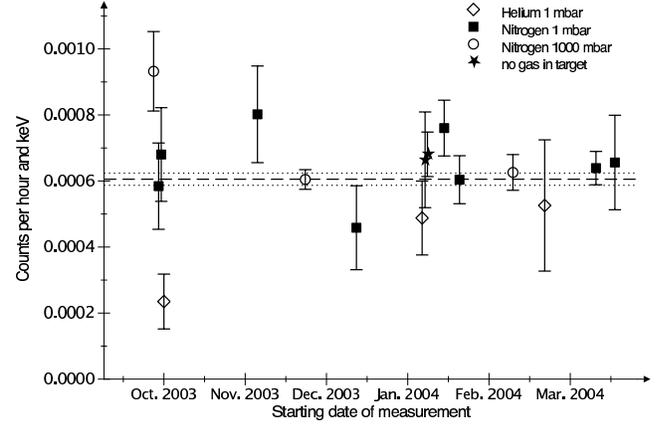}
 \caption{Laboratory background counting rate in the BGO detector of setup B in the \nuc{14}{N}(p,$\gamma$)\nuc{15}{O} region of interest over the period of the experiment for different gases inside the target chamber.}
 \label{NaturalBG-Countingrate3.pdf}
\end{figure}

In order to evaluate the impact of a particular shielding, it is useful to determine the counting rates without ion beam in an energy interval where capture $\gamma$ rays are expected. 
Such values are listed in table \ref{NaturalBG-Countingrate-Ge-table} for the germanium detector and in table \ref{NaturalBG-Countingrate-BGO-table} for the BGO detector. The BGO background in the \nuc{14}{N}(p,$\gamma$)\nuc{15}{O} region of interest has been monitored over a period of six months with repeated measurements during accelerator down times. The results are displayed in fig. \ref{NaturalBG-Countingrate3.pdf} and show that the counting rate is stable and independent of the type and pressure of gas present in the target chamber. The weighted average of all background runs and its uncertainty are indicated by dashed lines in the figure.\par

% ============ NaturalBG-Countingrate3.pdf

%%%%%%%%%%%%%%
\subsection{Background induced by the incident proton beam}
%%%%%%%%%%%%%%

While the laboratory background can be reduced by proper shielding, it is difficult and in some cases impossible to shield the detector against $\gamma$ rays arising from parasitic reactions induced by the ion beam incident on the target system. Previously to the LUNA solid target study of the \nuc{14}{N}(p,$\gamma$)\nuc{15}{O} reaction, the proton beam induced background for a setup equivalent to setup A has been investigated in the energy region from $E_{\rm p}$\,=\,140\,-\,400\,keV \cite{Strieder03-Debrecen}. It was found that the principal background reactions were 
\nuc{11}{B}(p,$\gamma$)\nuc{12}{C}, 
\nuc{18}{O}(p,$\gamma$)\nuc{19}{F}, and 
\nuc{19}{F}(p,$\alpha \gamma$)\nuc{16}{O}. 
They originated from the target itself, and a reduction in their yield was achieved by making adjustments in target production and preparation. The following discussion is therefore limited to the gas target setups B and C.\par

\begin{figure}[bt]
 \centering 
 \includegraphics[width=85mm]{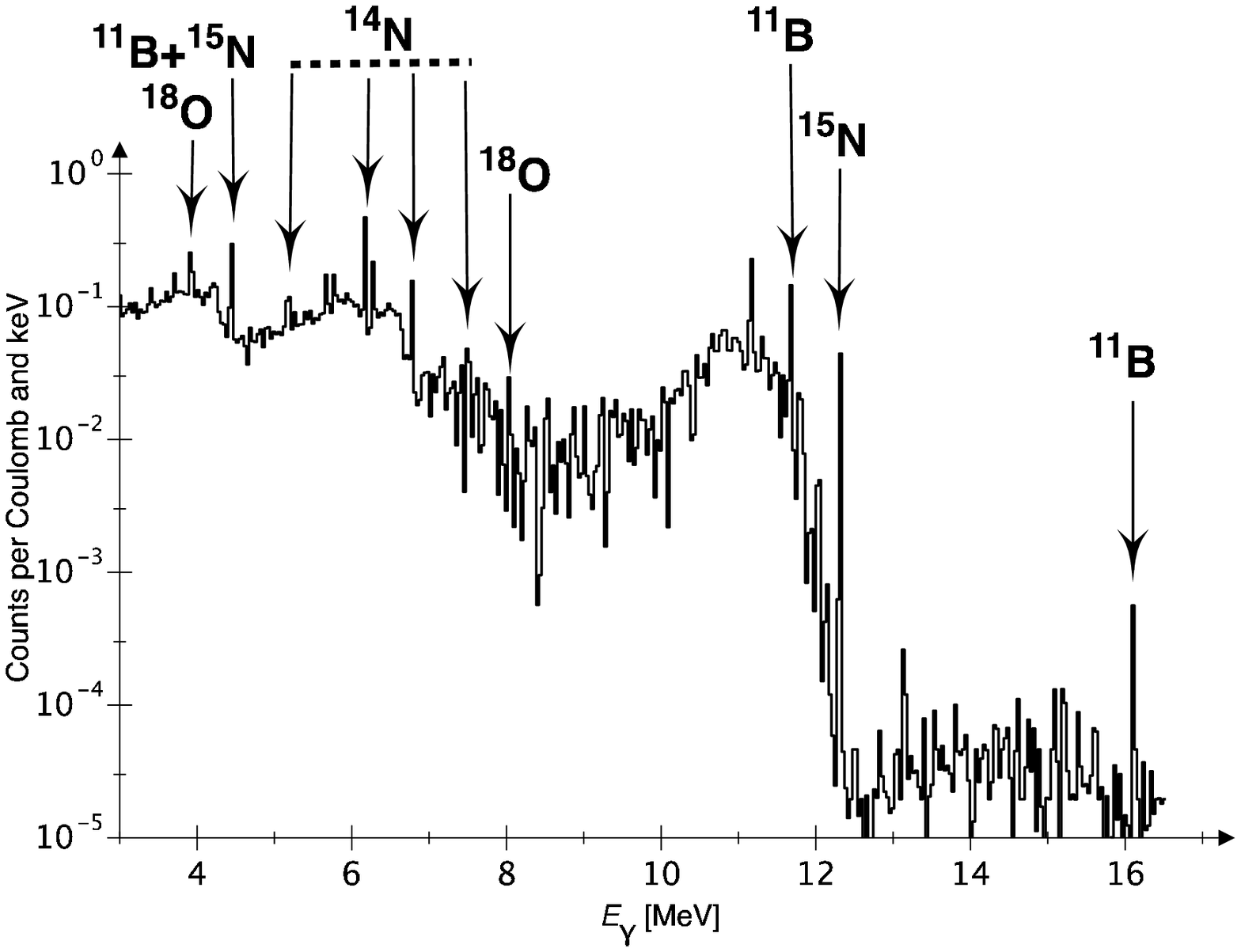}
 \includegraphics[width=85mm]{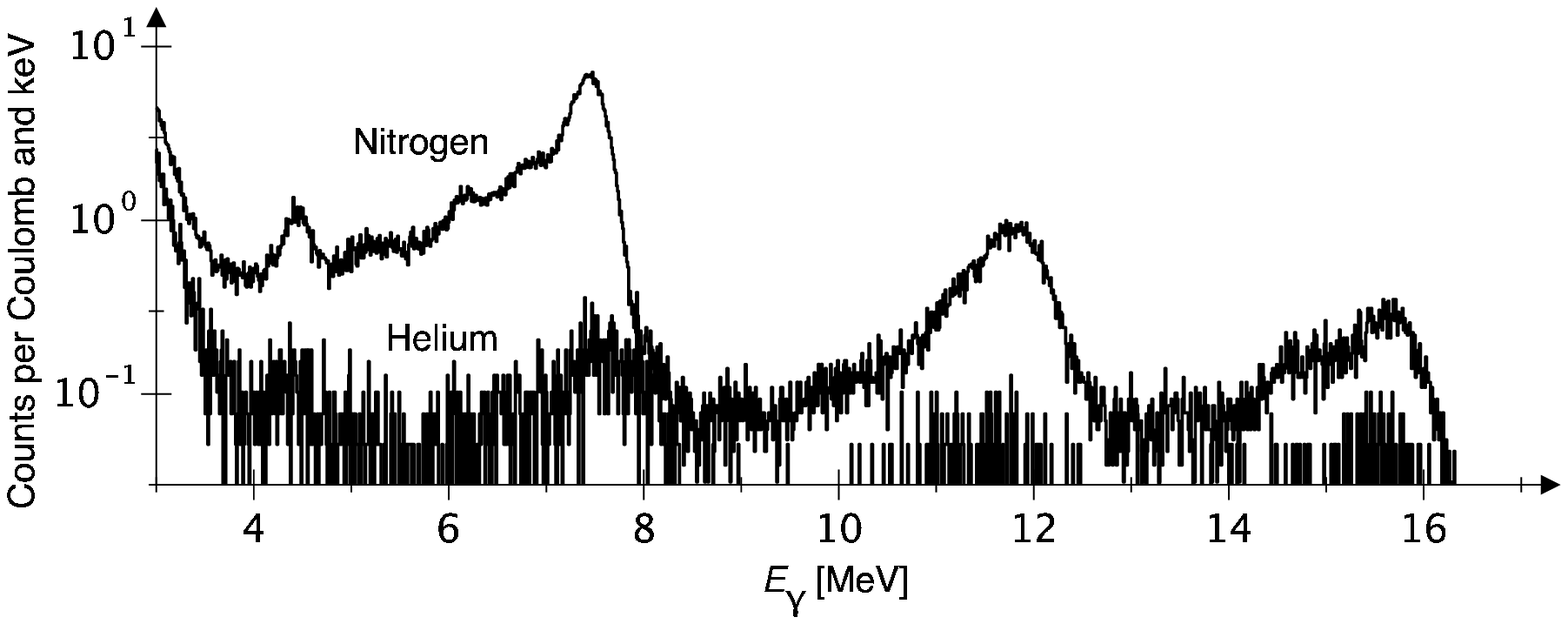}
 \caption{Spectra for $E_{\rm p}$\,=\,200\,keV with 1\,mbar gas in the target. -- Top panel: Germanium detector, setup C, nitrogen gas. -- Bottom panel: BGO detector, setup B, one run with nitrogen gas and one run with helium gas.}
 \label{Spec-200.pdf}
\end{figure}

%Daniel 15 Jan 05 begin
Several monitor runs were performed between $E_{\rm p}$\,=100 and 370\,keV with setup C. 
The beam current was typically 250\,$\mu$A, with 1\,-\,2\,\% of the target current being lost on the final aperture ($d$\,=\,7\,mm). 
%Daniel 15 Jan 05 end
A spectrum obtained with the germanium detector at $E_{\rm p}$\,=\,200\,keV proton energy and with 1\,mbar nitrogen as target gas is shown at the top of fig. \ref{Spec-200.pdf}. 
In the spectrum, the most important background lines as well as the lines from the \nuc{14}{N}(p,$\gamma$)\nuc{15}{O} reaction are identified. The \nuc{15}{N}(p,$\gamma$)\nuc{16}{O} and \nuc{15}{N}(p,$\alpha \gamma$)\nuc{12}{C} background results from the natural isotopic composition of the target gas (0.4\,\% \nuc{15}{N}). 
The \nuc{2}{H}(p,$\gamma$)\nuc{3}{He} and \linebreak
\nuc{13}{C}(p,$\gamma$)\nuc{14}{N} background lines are much less intensive than the \nuc{14}{N}(p,$\gamma$)\nuc{15}{O} lines at this beam energy and therefore not visible in the figure, but they gain in relative importance with decreasing beam energy. 
The \nuc{18}{O}(p,$\gamma$)\nuc{19}{F} and the \nuc{19}{F}(p,$\alpha \gamma$)\nuc{16}{O} reactions play an important role for runs close to their resonance energies at $E_{\rm p}$\,=\,151 and 224\,keV, respectively. 
%Daniel 15 Jan 05 begin
In runs with beam energies more than 20\,keV away from one of these resonances, the contribution of the corresponding reaction was found to be much smaller than that of the \nuc{14}{N}(p,$\gamma$)\nuc{15}{O} reaction to be studied.%
%Daniel 15 Jan 05 end
\par

% ============ Spec-200.pdf

At the bottom of fig. \ref{Spec-200.pdf}, two spectra obtained with the BGO detector of setup B at $E_{\rm p}$\,=\,200\,keV are shown: one denoted as 'nitrogen' with 1\,mbar nitrogen as target gas, the other denoted as 'helium' with 1\,mbar helium as target gas. It is clear that the high resolution germanium detector is needed in order to identify the background visible in the spectra taken with the BGO detector. The lines at $E_\gamma$\,=\,4.4 and 12\,MeV evident in both BGO spectra are from reactions on \nuc{11}{B} and \nuc{15}{N}. The line at 16\,MeV is due to the \nuc{11}{B}(p,$\gamma$)\nuc{12}{C} reaction caused by a \nuc{11}{B} impurity on the final collimator. The fact that it is weaker in the helium spectrum than in the nitrogen spectrum is due to better focusing of the beam for that particular helium run. At 7.7\,MeV in the helium spectrum, there is a line due to the \nuc{13}{C}(p,$\gamma$)\nuc{14}{N} reaction which is not visible in the nitrogen spectrum, because it is buried under the 7.5\,MeV sum peak from \nuc{14}{N}(p,$\gamma$)\nuc{15}{O}. 
%Daniel 15 Jan 05 begin
The small structure at 8.2\,MeV in both BGO spectra is due to the \nuc{18}{O}(p,$\gamma$)\nuc{19}{F} reaction.
%Daniel 15 Jan 05 end
The peaks at 5.2, 6.2 and 6.8\,MeV in the nitrogen spectrum are due to \nuc{14}{N}, with possible contributions from \nuc{2}{H} at 5.6\,MeV and \nuc{19}{F} at 6.1\,MeV. In some BGO spectra not shown here, there is a small feature above 14\,MeV that is attributed to the \nuc{7}{Li}(p,$\gamma$)\nuc{8}{Be} reaction.\par

% =========================================================
\subsection{Localisation of ion beam induced background using the Doppler shift}
% =========================================================

In order to understand the $\gamma$ ray background in a radiative capture experiment, it is necessary to identify the background reaction. This can be achieved with a germanium detector. Spatial information can then be extracted using the Doppler shift of the $\gamma$ lines. For a given transition and beam energy, the sign and magnitude of this shift depend only on the angle of emission $\theta$, as measured from the beam direction, allowing to localize the source of the $\gamma$ rays \cite{Bemmerer04-Diss}. \par

% ================== d + p Figure
\begin{figure}[t]
 \centering 
 \includegraphics[width=85mm]{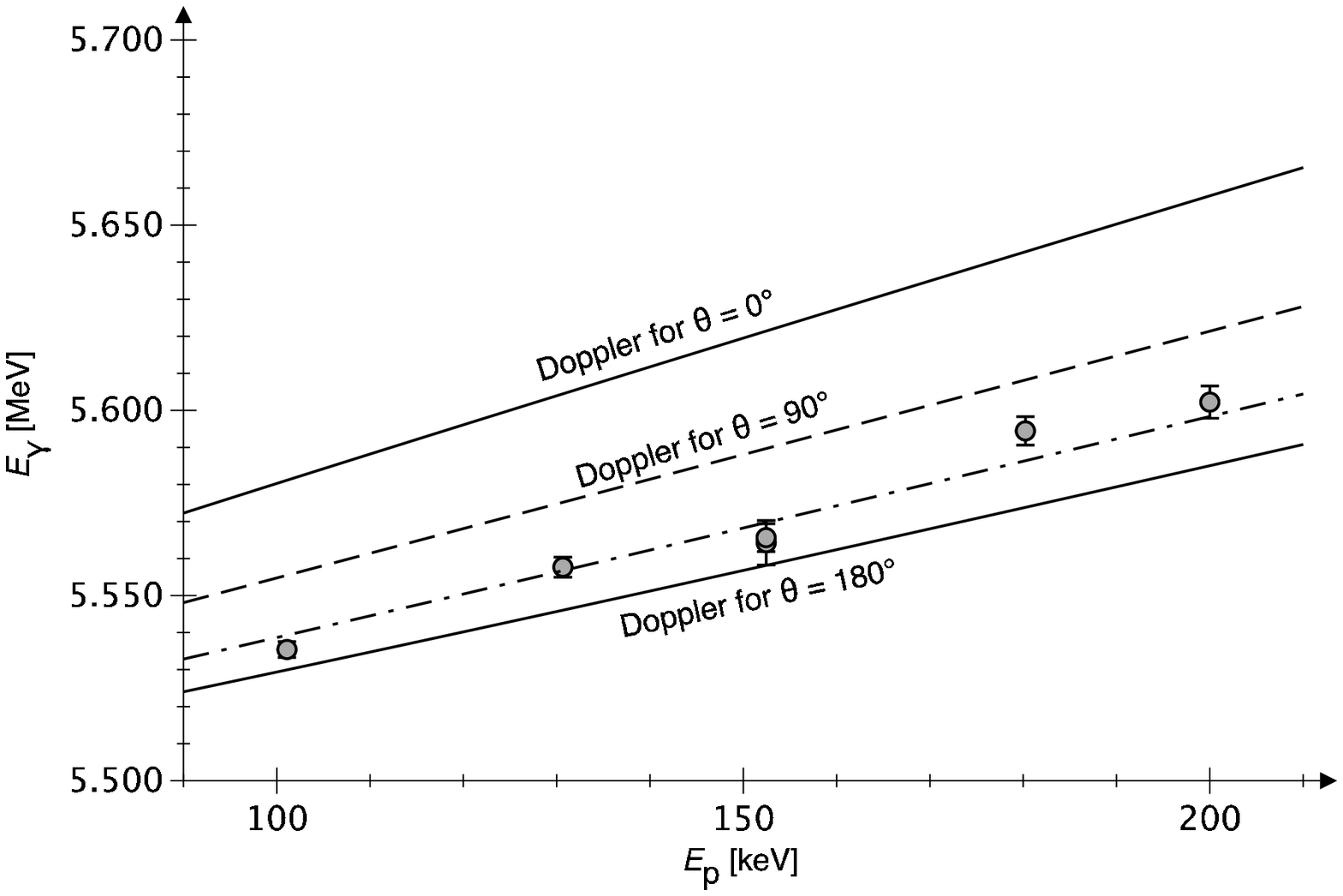}
 \includegraphics[width=85mm]{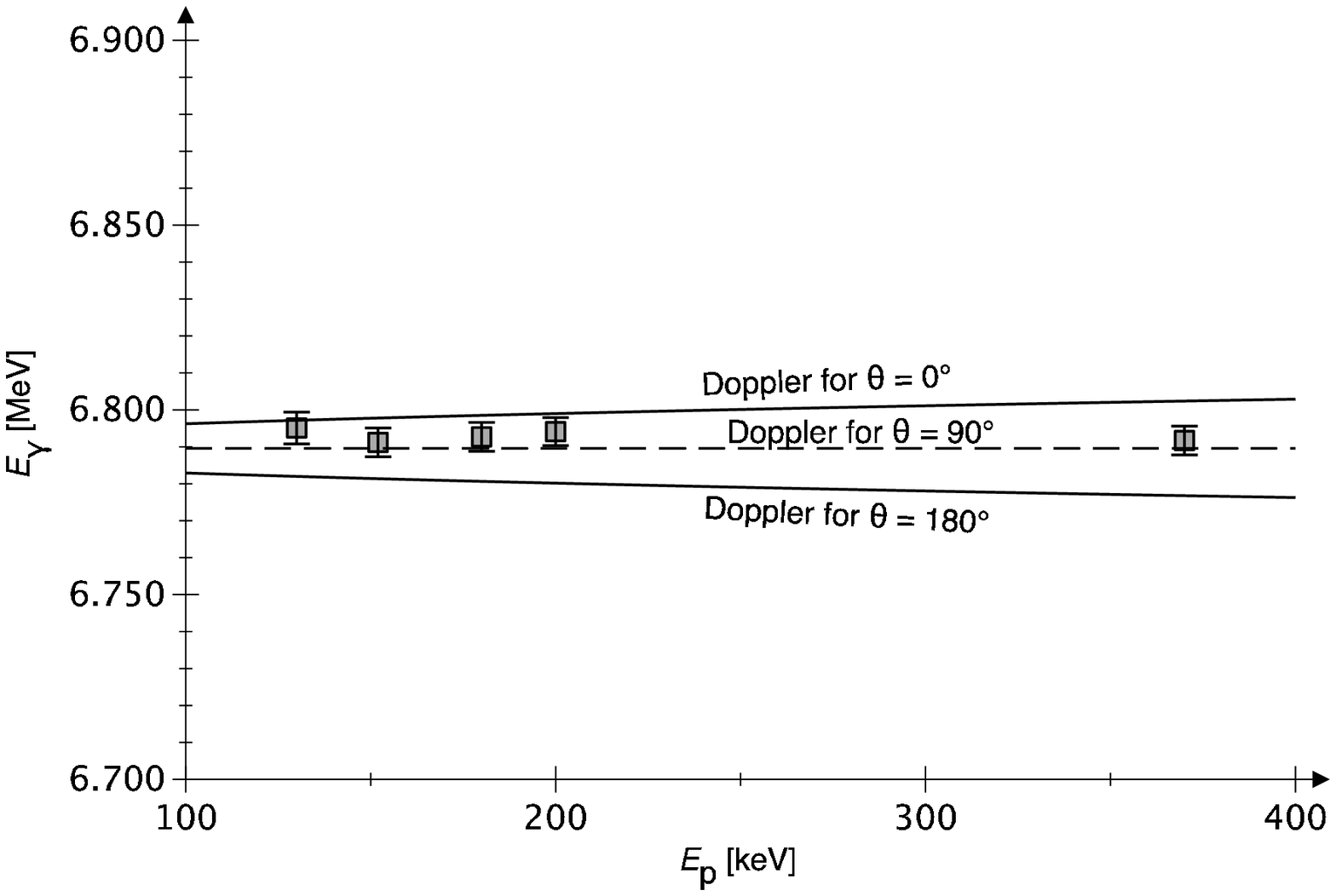}
  \caption{Top: Energy of the \nuc{2}{H}(p,$\gamma$)\nuc{3}{He} direct capture line plotted as a function of proton beam energy. -- Bottom: Analogous plot for the \nuc{14}{N}(p,$\gamma$)\nuc{15}{O} 6.79\,MeV secondary $\gamma$ line.}
 \label{d+p-Doppler}
\end{figure}

Such an experiment was performed at beam energies $E_{\rm p}$\,=\,100\,-\,370\,keV, using setup C. The proton beam hits only the final collimator and the beam stop, and thus beam induced $\gamma$ rays not coming from the target gas are only expected from forward angles (collimator, $\theta$\,$<$\,90\dgs) or backward angles (beam stop, $\theta$\,$>$\,90\dgs).
The measured $\gamma$ energies for the \nuc{2}{H}(p,$\gamma$)\nuc{3}{He} background reaction are displayed at the top of fig. \ref{d+p-Doppler} as a function of beam energy,
%Daniel 15 Jan 05
with the error bars reflecting both the statistical uncertainties and the systematic uncertainty stemming from the energy calibration of each spectrum.
%Daniel 15 Jan 05
The calculated $\gamma$ energies, with the proper Doppler and recoil corrections applied, are plotted as lines in the figure. The two solid lines are calculated with the Doppler correction for an angle of $\theta$\,=\,0\dgs\ and 180\dgs\ between emission and beam direction, respectively. The dashed line is for $\theta$\,=\,90\dgs, resulting in a zero Doppler correction. The dot-dashed line is a fit to the experimental points with $\theta$ as fit parameter, resulting in $\theta$\,$\approx$\,130\dgs. Only a backward angle, $\theta$\,$>$\,90\dgs, is compatible with the data. \par

In order to test the localization by the Doppler shift, an analogous graph is shown at the bottom of fig. \ref{d+p-Doppler} for the most intense $\gamma$ ray from \nuc{14}{N}(p,$\gamma$)\nuc{15}{O}, i.e. the transition from the 6.79\,MeV state to the ground state. The expected Doppler effect, denoted by solid lines in the picture, is smaller here because of the larger mass of \nuc{15}{O} when compared to \nuc{3}{He}, and there is no slope for the $\theta$\,=\,90\dgs\ curve, because it is a secondary transition. Since the nitrogen gas is mainly in front of the detector, one expects on average $\theta$\,$\approx$\,90\dgs, in good agreement with the data.\par

The same procedure has been applied to several other background reactions, using primary or secondary $\gamma$ rays. The $\gamma$ energies used and the results are summarized in table \ref{BeamIndBG-Reactions-table}. Knowing the location of origin made it possible to take steps that reduced the impact of the listed background reactions. For example, the final collimator, which was made from brass, was covered with a copper disk, resulting in a visible reduction in the \nuc{11}{B} background $\gamma$ rays.\par

% == BeamIndBG-Reactions-table
\begin{table}[tb] 
 \centering
\begin{tabular}{|c|r|l|c|}
\hline
\bf Reaction \raisebox{3mm}{} & \boldmath\bf $Q$ [MeV] & \boldmath\bf $E_\gamma$ [MeV] & \boldmath $\theta$ \\ \hline 
\raisebox{3mm}{}\nuc{2}{H}(p,$\gamma$)\nuc{3}{He} & 5.493 & 5.493+$E$ & $>$ 90\dgs   \\  
\nuc{11}{B}(p,$\gamma$)\nuc{12}{C} & 15.957 & 4.439 & $<$ 90\dgs\\ 
\nuc{13}{C}(p,$\gamma$)\nuc{14}{N} & 7.551 & 7.551+$E$ & $>$ 90\dgs  \\ 
\hline
\end{tabular}
 \caption{Sources of beam induced background in setup C, with the angle $\theta$ determined by the Doppler shift. Angles $\theta$\,$<$\,90\dgs\ correspond to the final collimator, and $\theta$\,$>$\,90\dgs\ to the beam stop.}
 \label{BeamIndBG-Reactions-table}
\end{table}

% =========================================================
\section{Feasibility of radiative capture experiments at the LUNA facility} 
% =========================================================

\begin{figure}[bt]
 \centering 
 \includegraphics[width=85mm]{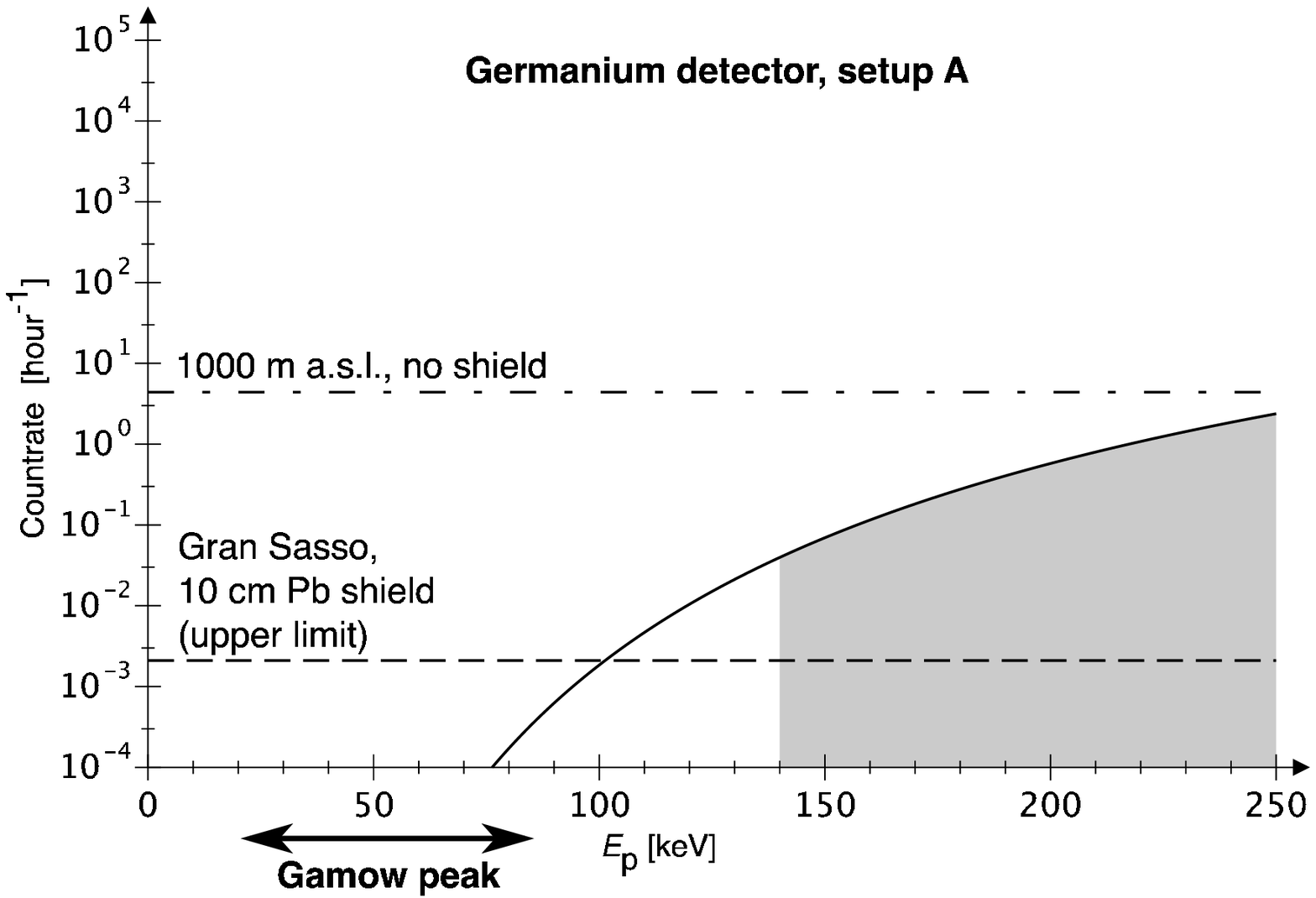}
 \includegraphics[width=85mm]{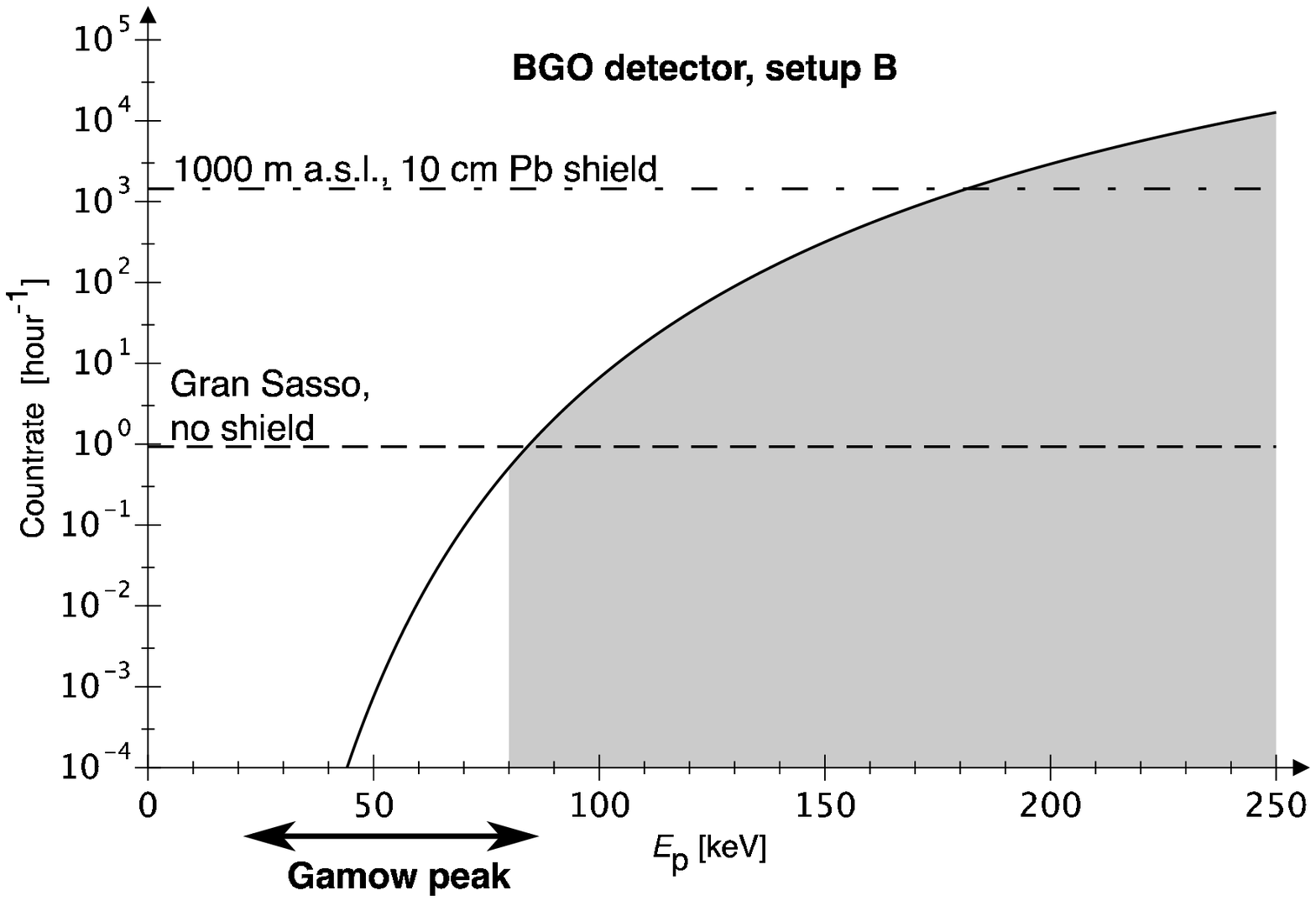}
  \caption{Predicted counting rate (solid curve) from the \nuc{14}{N}(p,$\gamma$)\nuc{15}{O} reaction, compared with background rates without ion beam. -- Top: Transition to the ground state in \nuc{15}{O} with a germanium detector and setup A. -- Bottom: Total cross section with BGO detector and setup B.}
 \label{Sensitivity.pdf}
\end{figure}

For two different scenarios at the LUNA2 400\,kV accelerator, expected counting rates in \nuc{14}{N}(p,$\gamma$)\nuc{15}{O} experiments have been calculated, 
assuming a constant S factor equal to the value for $E_{\rm p}$\,=\,150\,keV taken from \cite{Formicola04-Diss,Bemmerer04-Diss}, 
a beam current of 250\,$\mu$A, and the known detection efficiencies. The expected rates (solid lines in fig. \ref{Sensitivity.pdf}) can be compared to the background rates present without ion beam, as taken from tables \ref{NaturalBG-Countingrate-Ge-table} and \ref{NaturalBG-Countingrate-BGO-table}.\par

% ================== Sensitivity.pdf

In the top panel of fig. \ref{Sensitivity.pdf}, such a comparison is shown for the $\gamma$ ray from capture into the ground state of \nuc{15}{O}, using the germanium detector of setup A in close geometry ($d$\,=\,1.5\,cm) to a TiN solid target of 10\,keV thickness. The bottom panel of the figure shows the same plot for the total S factor and the BGO detector of setup B, with the target chamber filled with 1\,mbar nitrogen gas, corresponding again to about 10\,keV target thickness. In both figures the low energy part of the region under study at the LUNA facility in a corresponding setup is shaded. The energy of the Gamow peak is marked for a star in a globular cluster that is at the evolution stage used for the cluster age determination.
The objective of the LUNA solid target experiment was to clarify the contribution of the transition to the ground state in \nuc{15}{O}. Data were limited to energies above the laboratory background, because the sizable summing correction in close geometry made it impossible to obtain further information on this transition with acceptable uncertainty \cite{Formicola04-PhysLettB}. \par

For both LUNA \nuc{14}{N}(p,$\gamma$)\nuc{15}{O} experiments between \linebreak $E_{\rm p}$\,=\,80 and 400\,keV, the proton beam induced background was not the limiting factor for the sensitivity, although it contributed to the uncertainty. Because this background strongly depends on the particularities of the setup, on the projectile, and on the beam energy, no general statement on beam induced background for a given capture reaction can be made. \par

% =========================================================
\section{Summary and outlook}
% =========================================================

The special features of the LUNA facility have been reviewed, focusing on aspects important for radiative capture experiments 
giving rise to $\gamma$ rays with energies above 3.7\,MeV.
The \nuc{2}{H}(p,$\gamma$)\nuc{3}{He}, \nuc{14}{N}(p,$\gamma$)\nuc{15}{O}, and \linebreak \nuc{25}{Mg}(p,$\gamma$)\nuc{26}{Al} reactions were used as examples. 
The laboratory gamma background has been investigated using both germanium and BGO detectors. For the example of the LUNA gas target setup, the proton beam induced background has been discussed. The sources of beam induced background have been localized using the Doppler effect. Based on the present data, the feasibility of radiative capture experiments for nuclear astrophysics has been evaluated in different shielding scenarios.\par 

In order to study reactions with $\gamma$ energies lower than 3.7\,MeV, a shielding similar to that of low level counting facilities would be required, with a thick lead shield complemented by an inner copper lining. Such a setup is currently under construction for a future LUNA \nuc{3}{He}($\alpha$,$\gamma$)\nuc{7}{Be} ($Q$\,=\,1.59\,MeV) experiment \cite{LUNA-AR2003}.\par

In conclusion, the installation of accelerators at the Gran Sasso underground laboratory with its effective cosmic ray shield allows to measure the cross sections of astrophysically relevant reactions at energies that are much lower than those accessible in laboratories at the earth's surface. In many cases, one can even reach the Gamow peak for important stellar scenarios. \par

% =========================================================
\section*{Acknowledgments}

This work was supported in part by: INFN, TARI HPRI-CT-2001-00149, OTKA T 42733 and T 49245, BMBF (05CL1PC1-1), and FEDER-POCTI/FNU/41097/2001. 

% =========================================================
%\bibliography{/Users/daniel/Library/TeX/Danielsbib} 
%\bibliographystyle{h-physrev3} 

% =========================================================
\end{document}